This manuscript has been authored by UT-Battelle, LLC, under Contract No. DE-AC0500OR22725 with the U.S. Department of Energy. The United States Government retains and the publisher, by accepting the article for publication, acknowledges that the United States Government retains a non-exclusive, paid-up, irrevocable, world-wide license to publish or reproduce the published form of this manuscript, or allow others to do so, for the United States Government purposes. The Department of Energy will provide public access to these results of federally sponsored research in accordance with the DOE Public Access Plan (http://energy.gov/downloads/doe-public-access-plan).

# Deep data mining in a real space: Separation of intertwined electronic responses in a lightly-doped BaFe$_2$As$_2$

Maxim Ziatdinov[†‡], Artem Maksov[†§], Li Li[∥], Athena S. Sefat[∥]

Petro Maksymovych[†‡], and Sergei V. Kalinin[†‡§]

[†]*Center for Nanophase Materials Sciences, Oak Ridge National Laboratory,*

*Oak Ridge, TN, 37831*

[‡] *ORNL Institute for Functional Imaging of Materials, Oak Ridge National Laboratory,*

*Oak Ridge, TN, 37831*

[§]*Bredesen Center for Interdisciplinary Research, University of Tennessee,*

*Knoxville, TN 37996*

[∥]*Material Science & Technology Division, Oak Ridge National Laboratory,*

*Oak Ridge, TN, 37831*

To whom correspondence should be addressed:

E-mail: ziatdinovma@ornl.gov

E-mail: sergei2@ornl.gov

**Abstract**

**Electronic interactions present in material compositions close to the superconducting dome play a key role in the manifestation of high-$T_c$ superconductivity. In many correlated electron systems, however, the parent or underdoped states exhibit strongly inhomogeneous electronic landscape at the nanoscale that may be associated with competing, coexisting, or intertwined chemical disorder, strain, magnetic, and structural order parameters. Here we demonstrate an approach based on a combination of scanning tunneling microscopy/spectroscopy (STM/S) and advanced statistical learning for an automatic separation and extraction of statistically significant electronic behaviors in the spin density wave (SDW) regime of a lightly (~1%) gold-doped $BaFe_2As_2$. We show that the decomposed STS spectral features have a direct relevance to fundamental physical properties of the system, such as SDW-induced gap, pseudogap-like state, and impurity resonance states.**

**Introduction**

Nanoscale inhomogeneity of chemical, structural and electronic orders in a crystalline matter is expected to have a profound and non-random effect on the macroscopic properties of technologically relevant materials. Celebrated examples include reduced mobility of Dirac electrons in graphene transistor devices due to formation of charge nanopuddles,[1,2] ultra-high piezoelectric response of relaxor ferroelectrics due to interaction between nanopolar domains and acoustic phonon mode,[3] filamentary superconductivity,[4] and fluctuating superconducting (SC) state above a transition temperature ($T_c$) in high-$T_c$ cuprates associated with emergence of nanometre-sized electron pairing regions.[5]

Scanning tunneling microscopy and spectroscopy (STM/S), which probes topographic and electronic properties of the surfaces with a nanometer-scale resolution, constitutes an ideal experimental tool for exploring local inhomogeneity in materials. The STM topographic images are typically recorded in a constant current regime,[6] resulting in a 2-dimensional (2D) $Z(X,Y)$ dataset, where $Z$ represents a convolution of actual height variation and electronic local density of states (LDOS) at each point $(X,Y)$ on the surface. Meanwhile, the STS mode allows to acquire 3D $G(X,Y,V)$ datasets, where $G=dI/dV$ corresponds to a value of differential conductance proportional to LDOS at specific energy $E=eV$ at each $(X,Y)$ point. In the simplest realizations of the bi-phase or multi-phase nanoscale systems, a separation between two or more phases is clearly visible in the STM topography, and comparison of STS spectra associated with different topographic features allows a straightforward analysis of electronic properties in these phases. Examples include STM/S measured on 2D lateral heterostructures sufficiently far from the boundary[7] or STM/S experiments on an isolated impurity embedded in otherwise ideal lattice[8]. For many strongly correlated materials, however, a complex local inhomogeneity patterns in conductance maps do not have a direct and simple connection to topographic features.[9-11] To complicate things even further, the morphology and chemical composition of the top-most layer of a cleaved surface in many complex compounds is usually itself a subject of controversy[12] which makes it nearly impossible to predict electronic properties in a characteristic field of view (FOV) of STM/S experiment from the first principles.

Given an ever-growing amount of multidimensional STM/S data on strongly correlated materials,[12-14] there is an urgent need for developing a deep data driven analysis that would allow reliable and un-biased identification and spatial mapping of statistically significant different electronic behaviors without *a priori* knowledge about the details of surface structure. Here we present a physics-robust statistical learning approach based on *k*-means clustering, principle component analysis (PCA) and Bayesian linear unmixing to uncover a wealth of "hidden" information from the STS measurements in a lightly-doped, "precursor", magnetic regime of iron-based superconductor.[15] We show how the features extracted from multivariate statistics-based decomposition of STS signal have a direct relevance to fundamental physical properties of the system, which we illustrate by uncovering a "buried" pseudogap-like phase and impurity induced double resonance states.

As a model system, we have chosen a lightly Au-doped $BaFe_2As_2$ single crystal, $Ba(Fe_{1-x}Au_x)_2As_2$ with x=0.009 ( ~1%). This compound shows a coupled structural and antiferromagnetic (AF) transition, from the tetragonal non-magnetic state into the orthorhombic striped SDW phase at $T_N \approx 110$ K.[16] Upon increased Au-dopants, the AF interactions becomes suppressed and the system develops into a superconductor ($T_c \approx 4$ K) at ~3%. It has been recognized that interactions present in such SDW states of the FeAs-based compounds play a crucial role in understanding unconventional superconductivity.[17, 18] However, the details of local electronic structure at low temperatures in the non-SC phase of FeAs compounds, including the role of lattice strain, presence and origin of a pseudogap-like state, and character of impurity-induced quasiparticle states, remain a subject of a debate.

## Results and discussion

We first present the STM topographic image over a relatively large FOV on a cleaved surface of 1% Au-doped $BaFe_2As_2$ in the SDW phase recorded at T=77 K (Fig. 1a). The typical surface area at 77 K appears to be peppered with dark nanoscale regions. Upon cooling down to T=4 K, we found a dramatic increase in the density of the dark nanoregions as can clearly be seen from the representative STM topographic image for the same Au-doped $BaFe_2As_2$ surface in Fig. 1b. In general, the variations in apparent topographic height associated with dark and bright regions can be of both topographic and electronic origin. However, we do not expect any extensive surface damage or profound changes in nanoscale chemical composition as we cool down the sample from 77 K to 4 K. Instead, the observed change in STM topographic patterns in Fig. 1a and 1b is likely related to the enhanced nanoscale electronic inhomogeneity as we approach towards a phase region with competing normal and SC orders[19] or with admixture of another form of magnetic order within the SDW phase (See Supplemental Material[20]). Such inhomogeneity shows the necessity for applying data mining tools based on multivariate statistical analysis for extracting relevant electronic behaviors in this system[21].

Zooming into a smaller FOV reveals a stripe-like reconstruction at the surface with a periodicity across the stripes of ≈0.7 nm (inset in Fig. 1b). Similar unidirectional modulation of charge density has been also reported for SDW phase of $SrFe2As_2$.[22] While the exact origin of these charge stripes and their relation to SDW, if any, is not clear at present moment, it is worth to note that we were not able to observe similar 1D modulations at 77 K on the same cleaved surface. This suggests that the reconstruction is not cleavage-induced.

In Fig. 1c we show the STM topography at *T*=4 K measured in a region with extended quasi-1D defect which appears as a bright "diagonal" feature in the topography. The spatial extension of this defect typically exceeds ≈1 μm and we were able to reproducibly observe it in several areas of the sample. Furthermore, a similar structure was reported in another iron-based superconductor compound[23] suggesting that this defect can be a common feature of iron pnictides. High-resolution spatial maps of differential conductance $G$ (Fig. 1d-g) recorded at the area shown in Fig. 1c at several selective energies confirm a highly inhomogeneous electronic structure of the surface, with no one-to-one correspondence to underlying topographic data. Accordingly, averaging over even relatively small surface area can lead to a loss of significant physical information contained in individual $G$ curves. Such a lossy compression of the original data is illustrated in Fig. 2b, where the STS curve averaged from 182 individual line spectra inside the box in Fig. 2a fails to reproduce physically important features at the Fermi level seen in the 4 selected point spectra recorded within the box area. It therefore becomes clear that the surface electronic behavior cannot be characterized reliably by a simple visual assessment of the topographic image and individual inspection of STS curves from $G(X,Y,V)$ dataset.

We now proceed to the accurate extraction of statistically significant information associated with surface electronic structure from a deep data style analysis.[21] We use the STS dataset recorded on the topographic area shown in Fig. 2a. The dataset has dimensions of $X \times Y \times V = 80 \times 80 \times 768$, that is, it contains a stack of 768 conductance maps with a spatial resolution 80px×80px. To decorrelate the STS data in a statistically meaningful way we start with imposing a lower bound limit on the number of relevant electronic behaviors within the dataset. The smallest reasonable number of statistically significant behaviors can be estimated using $k$-means algorithm.[25] The $k$-means algorithm divides the dataset in a specified number of optimally selected clusters of curves that have similar behavior so that the within-cluster sum of squares is minimized:[24, 25]

$$\arg\min \sum_{i=1}^{k} \sum_{x_j \epsilon S_i} \|x_j - \mu_i\|,$$

where $\mu_i$ is the mean of points in $S_i$. The selection of the number of clusters is based on the analysis of dendogram in Fig. 3b, in which larger vertical drops in the binary branches indicate a better cluster organization scheme in the data[25]. Based on the results shown in Fig. 3b, we used 3 clusters as an input in our $k$-means analysis. The resultant spatial distribution of the 3 clusters is shown in Fig. 3a, and the mean curves associated with each of 3 clusters are displayed in Fig. 3c-e.

We further analyzed a variance in the STS curves distributed over each nanoregion (cluster) by means of PCA.[26-28] The deviation from the mean curve within each cluster associated with first

5 eigenvectors in PCA is shown by dashed lines in Fig. 4a-c, and the corresponding scree plots are depicted in Fig. 4d-f.The PCA analysis indicates that the cluster 2 and cluster 3 show a relatively moderate variance in the shape of the mean STS curve, allowing us to extract physical information from the curves. The STS curve from cluster 2 displays a metallic behavior and a dip at about 5–10 meV below the Fermi level. This is in a good agreement with an observation of the SDW gap centered at around -10 meV in the ARPES measurements of $BaFe_2As_2$.[29] We note that the theoretical modelling in ref. 29[29] and ref. 30[30] also showed that the SDW phase features a finite density of states at the Fermi level in the absence of the (coexisting) superconducting state, which is supported by our results. The mean STS curve from cluster 3 shows a metallic behavior and is somewhat similar to the curve from cluster 2, but with the center of the dip shifted to about -25 meV. We tentatively assign this behavior to the SDW phase whose characteristics were altered locally due to the presence of distortion nano-domains in crystalline lattice induced by the quasi-1D defect. Noteworthy, we did not observe similar lineshape in the regions far (>100 nm) from the defect in our experiments. The PCA-derived variance within cluster 2 and cluster 3 can be understood as relatively minor fluctuations of electronic response within a defined phase. The situation, however, is quite different for cluster 1. Here, a stronger variance in the shape of STS curves, especially in the regions close to the Fermi level, does not allow assigning any physically-defined phase. This suggests that the total number of relevant electronic behaviors is larger than estimated by the $k$-means method. However, additional, "hidden", electronic responses cannot be accurately revealed from the PCA eigenvectors, as they are constructed to be orthogonal and hence may not have a well-defined physical meaning.

To perform a more thorough and detailed separation of electronic behaviors in the STS dataset we adopt Bayesian linear unmixing (BLU) technique. This algorithm, developed by Dobigeon and co-workers[31], is used for separating linear mixtures of spectral sources under non-negativity and full additivity constraints that allows assignment of physical meaning to the shape of the end-member curves.[32] We assume that the registered STM current signal $g$ at each pixel $p$ can be approximated as a linear combination of the currents flowing through each of the available "channels", so that the latter can be represented by the endmembers,[33]

$$g_p = \sum_{r=1}^{R} m_r a_{p,r} + n_p$$

$$G = MA + N$$

where $m_r$ is the spectral measurement of an endmember $r$, $R$ is the total number of endmembers, $a_{p,r}$ is the abundance of endmember $r$ at pixel $p$, $n_p$ is a zero-mean Gaussian noise, and $G = [g_1, ..., g_P]$, $M = [m_1, ..., m_R]$, $A = [a_1, ..., a_P]$, $N = [n_1, ..., n_P]$. The model is imposed with non-negativity and full-additivity constraints of the abundance coefficients:

$$a_{p,r} \geq 0, \quad r = 1, ..., R$$

$$\sum_{1}^{R} a_{p,r} = 1$$

Additionally, the model is constrained to non-negativity of the endmember spectra,

$$m_r \geq 0, \quad r = 1, ..., R$$

and no assumption of the presence of pure pixels is made. The estimation of endmembers and abundances is performed under a hierarchical Bayesian model.[31] For the estimation of the prior for the Bayesian model, the data $X = MA$ dimensionality is reduced to $K$ ($R - 1 \leq K \leq V$), by an assumption that without the noise data can be represented by ($R - 1$)-dimensional convex polytope of $R^V$, where vertices represent pure endmember spectra $m_r$. For the next step, PCA projection is obtained, which allows to recover a simplex in a reduced dimensionality space via NFINDR.[34,35] Using these results, the endmember abundance priors as well as noise variance priors are estimated from the conjugate multivariate Gaussian distribution, where the posterior distribution is calculated based on the endmember independence using Markov chain Monte Carlo (MCMC), which generates asymptotically distributed samples probed via Gibbs sampling strategy. Unmixing was run for 100 MCMC iterations for each attempt.

The number of endmembers $R$ in the BLU algorithm must be postulated by a researcher. The lower bound limit for a total number of endmembers has already been set by the results of $k$-means method. To set up the upper bound limit for possible number of relevant electronic behaviors, we refer to a general underlying physics of the problem. Here, in addition to the states associated with the SDW phase discussed in the $k$-means calculations, we must add states associated with (i) unidirectional modulation of surface charge density; (ii) possible presence of a different magnetic order "admixed" into the SDW phase; (iii) 2 common types of point defects on the cleaved surface; (iv) diluted concentration of Au dopants; (v) randomly scattered atoms on the surface.[20,36] Using these constraints, coupled with the results of PCA analysis, and by performing over- and under-sampling of BLU $R$-components, we found that the most relevant description of electronic behavior is achieved for $R=6$ endmembers.[20] The BLU results with $R=6$, for both endmembers and abundance maps, are shown in Fig. 5. One can immediately see that endmember 4 and endmember 5 (Fig. 5d and 5e) corroborate the results on SDW-associated phase found earlier from $k$-means algorithm. In addition to phases already seen in the $k$-means, the BLU analysis revealed new features in electronic behavior that can be linked to the fundamental physical properties of the material, as described below.

The endmember 2 shows a well-defined signature of a spectral gap of $2\Delta \approx 40$ meV centered near the Fermi level (Fig. 5b). We note that the gap of a similar behavior and magnitude (~30-40 meV) was observed in STM experiments on a closely related compound from $A$Fe$_2$As$_2$ family, SrFe$_2$As$_2$, in which it was explained as the SDW-originated gap.[22] However, recent photoemission spectroscopy measurements and theoretical modelling[29] on the $A$Fe$_2$As$_2$ type compound revealed that the SDW and SC orders must coexist spatially in order to produce a gap at the Fermi level. Otherwise, the SDW opens a gap below the Fermi level,[29,30] in agreement with behavior observed in the endmembers 4 and 5. If the SDW and SC orders indeed coexist on a local scale in our sample, the formation of the electron-pairing "islands" associated with the SC order is expected to produce a continuous drop in bulk resistivity measurements.[4, 37] However, this is clearly not the case for our compound, in which the resistance showed a small *upward* trend in the temperature range of interest.[20] This rules out a scenario in which the "admixture" of SC order leads to the gap feature seen in the endmember 2. We therefore describe the spectral weight loss at the Fermi level observed in the endmember 2 in terms of a pseudogap-*like* state, which is defined here as a state outside the "superconducting dome" and not directly associated with either SDW-induced gap or 'SDW+SC'-induced gap.

We next discuss a possible physical origin of the pseudogap-like state associated with endmember 2. At first glance, it is tempting to link the pseudogap-like state to the 1D striped charge order seen in the STM topographic image. However, a possibility of such direct correlation quickly falls apart as we were able to find the 1D stripes even in the areas without spectral features of a pseudogap. Another explanation of a pseudogap-like state is based on the possible formation of a different, short-range, magnetic order admixed into the SDW phase, which is consistent with the presence of upturn in a magnetic susceptibility data below $T_N$. The formation of a pseudogap-like state may also explain a peculiar upswing in the resistivity in the SDW phase below ≈20 K.[20]. Noteworthy, our finding of a $2\Delta$≈40 meV pseudogap-like feature, which is not directly related to SDW, correlates with photoemission spectroscopy results of Xu *et al*.[38] that showed emergence of $2\Delta$≈36 meV pseudogap state at the Fermi level of the underdoped $Ba_{1-x}K_xFe_2As_2$ in both SC phase and non-SC phase without long-range SDW order.

The endmembers 1 and 6 in Fig. 5a and 5f show a clear resonance peak features associated with impurity induced bound states in the SDW phase of this compound. The impurity-induced nature of these peaks is further confirmed by the inspection of the corresponding abundance maps that show the peak features are generally constrained to a point-like areas on the surface (Fig. 5g,l). Of particular interest is the endmember 1 which can be described by a non-magnetic impurity-induced double resonance peak model studied in ref. 30.[30] We tentatively ascribe the origin of this spatially diluted double peak state to the Au dopants. Finally, the origin of endmember 3 is likely related to the minor instabilities ("noise") of the tunneling junction during the grid acquisition.

## Conclusions and outlook

Our results on the identification of a surface nanoscale electronic structure in the underdoped state of FeAs-based superconductor, by means of a deep data style analysis are important for providing clues to understand how the high-temperature superconductivity may emerge in these systems. First, while there is a growing evidence of the pseudogap state formation in FeAs compounds,[38,40-42] there is still an open debate on the relation of the pseudogap to the superconducting state and on the role of magnetic correlations in the formation of the pseudogap. Our revelation of "buried" pseudogap-like spectral features in the SDW phase, combined with results of magnetic susceptibility and resistivity measurements, suggests a potential link between a pseudogap state and weak or short-range magnetism within SDW phase. In future, the temperature-dependent STS measurements coupled with the presented here deep data analysis could provide additional details on a physical nature of this phase. Second, the real-space analysis of the electronic character of impurity-induced quasiparticle states found in the spectral unmixing of our data can be further used as a probe into the details of the strong correlations in the system. In this sense, it is natural to extend the deep data approach to the reciprocal space, which is commonly used to study quasiparticle interference pattern.[43] We expect that a nanoscale inhomogeneity in the electronic structure of the surface would produce spatially different scattering patterns at the same value of energy. The application of techniques such as sliding FFT combined with multivariate analysis[44] would allow hidden scattering patterns to be uncovered. Finally, we note that the presence of 1D charge modulation at the surface did not allow us to measure the atomic lattice constant in the regions close to the defects that showed peculiar changes in electronic behavior within the SDW phase. We do expect, however, that for the systems in which the atomic lattice can be resolved (i.e., no surface "reconstructions" occur), one can perform a direct data mining to correlate minute variations in atomic positions with the changes in spectral characteristics, such as the magnitude of SDW and/or SC gaps. As the ever-increasing amount of STM/S data on strongly correlated systems makes the individual inspection of datasets highly impractical and, in many cases, nearly impossible, the approach outlined here will present an ideal tool for an accurate mapping of locally inhomogeneous electronic structure on the surfaces in an automated fashion of a full information extraction.

**Methods:**

Single crystals of lightly Au-doped $BaFe_2As_2$ (x= 0.009, ~1%) were grown out of self-flux using a high-temperature solution-growth technique. STM/S measurements presented in the manuscript were carried using a Joule-Thomson scanning tunneling microscope (JT-STM, Specs, Berlin). Tungsten (W) STM tip was prepared by gentle field emission at a clean Ag(111) sample. The samples were cleaved *in situ* in the STM machine chamber at approximately 110 K. The samples were always measured first at 77 K, followed by cooling the system further down and performing measurements at 4 K. The representative figures of the surface at 77 K and at 4 K were selected from data obtained by routine shift of the scan frame over (2×2) $\mu m^2$ surface area, which ensured the overlap between areas imaged at different temperatures. The estimated lateral 'drift' upon cooling in the current experimental set-up is less than 1 μm. STS measurements were performed using a standard lock-in amplifier techniques, with a bias modulation between 2 mV and 5 mV. Matlab codes were used for data processing and statistical analysis.

**Acknowledgment:**

This work was supported by the U.S. Department of Energy (DOE), Office of Science, Basic Energy Sciences (BES), Materials Science and Engineering Division. Research was conducted at the Center for Nanophase Materials Sciences, which is a DOE Office of Science User Facility.

**FIGURES**

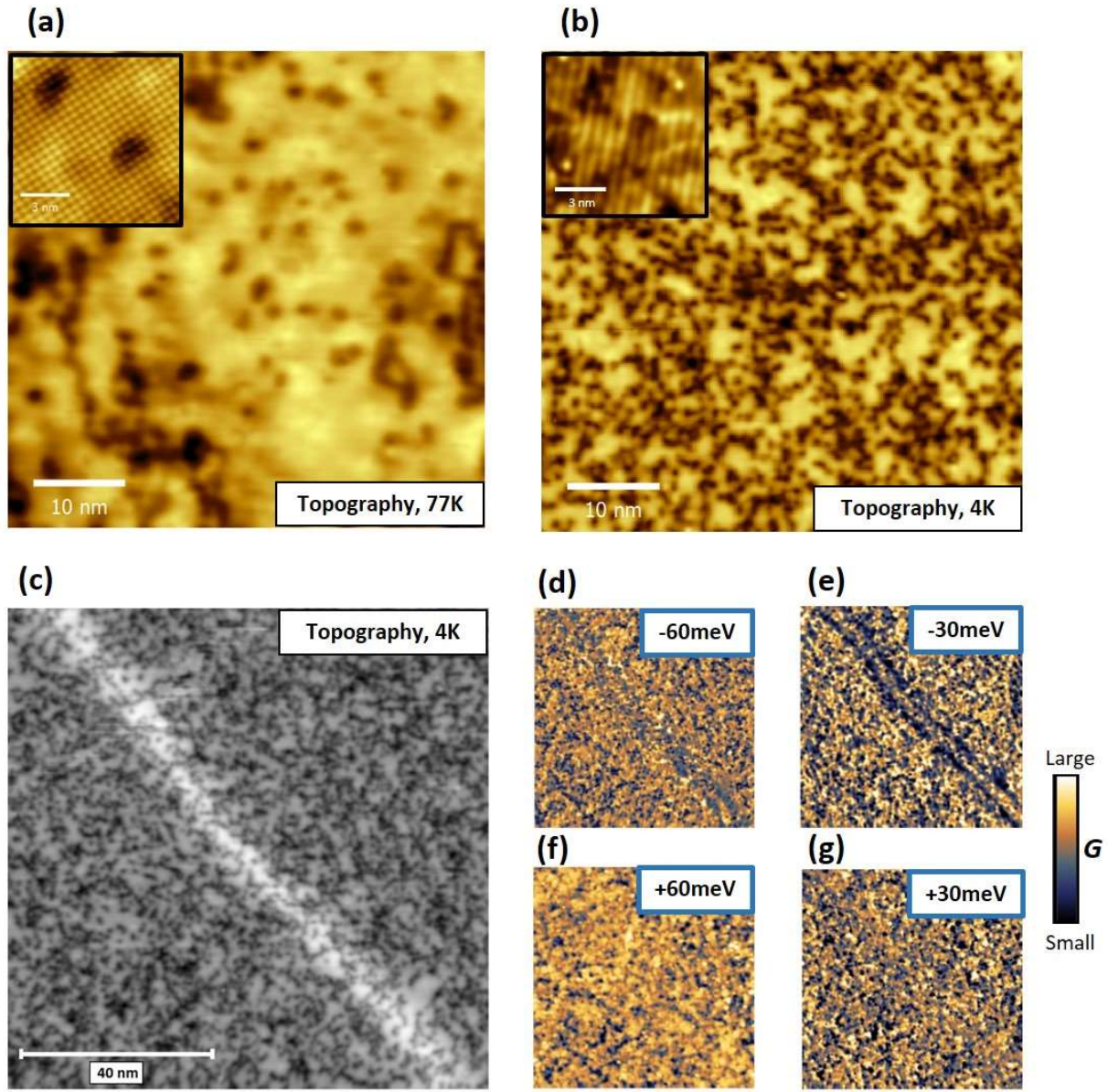

**FIGURE 1.** Data of 1% Au-doped $BaFe_2As_2$. (a,b) Representative experimental STM topographic images at 77K (a) and 4K (b). $U_{bias}$=-60mV, $I_{setpoint}$=100pA. Insets in (a) and (b) show a zoomed-in area of the surface which displays atomic lattice contrast ($U_{bias}$= -100 mV) and stripe-like features ($U_{bias}$=250 mV). (c) STM topographic image of quasi-1D defect at 4K. $U_{bias}$=-110mV, $I_{setpoint}$=150pA (d-g) Conductance maps $G$ $(r, E=eV)$ in the same area as in (c) at several selected energies.

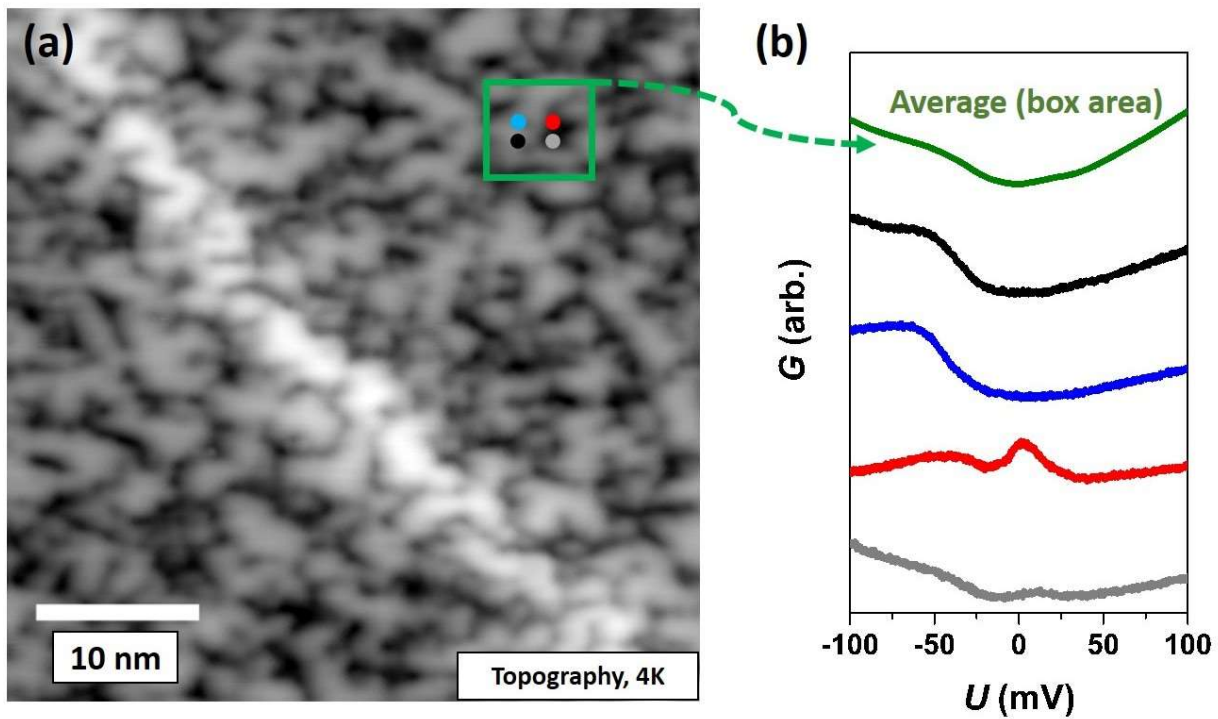

**FIGURE 2.** Data of 1% Au-doped $BaFe_2As_2$. (a) STM topographic image of the area on which STS grid measurements are performed. (b) Green STS curve is averaged over the 182 individual STS spectra inside the box in (a). Gray, red, blue, and black curve are single-point STS spectra recorded at the locations denoted by colored dots in (a). Note that gray and red curves were extracted from locations displaying similar topographic features.

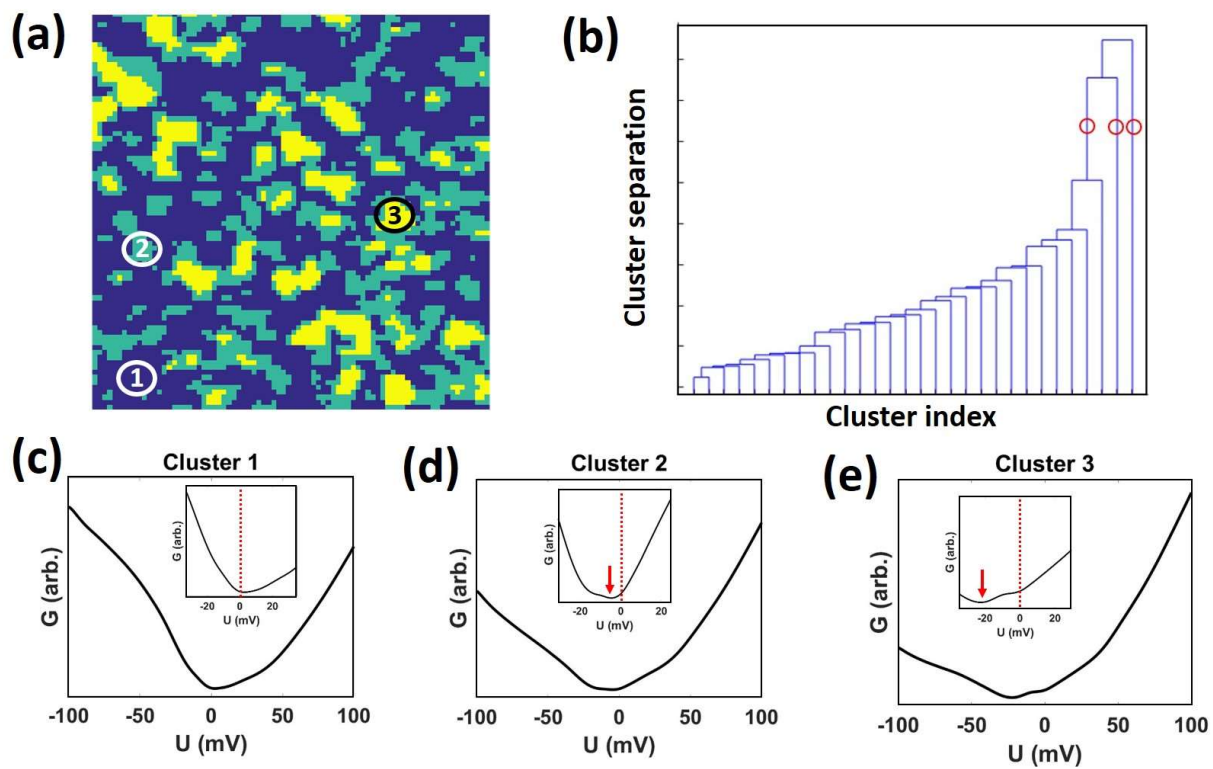

**FIGURE 3.** Results of *k*-means clustering. (a) *k*-means cluster algorithm resultant map with 3 clusters specified in the image. The surface area is identical to the one in Fig. 2(a). (b) Dendogram plot of hierarchical binary cluster tree (circles illustrate the optimal number of clusters). (c-e) Mean STS curves for each of 3 clusters. Insets show zoomed-in area near the Fermi level (dotted red line)

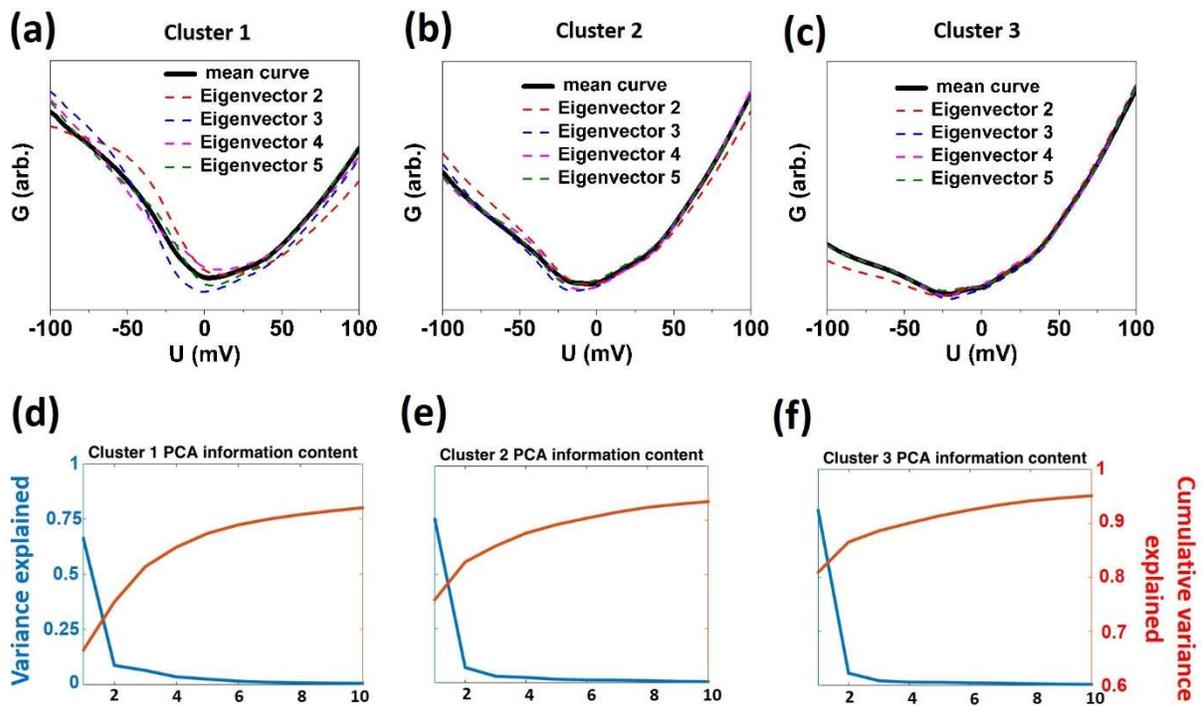

**FIGURE 4.** Results of PCA analysis performed within each of 3 clusters. (a-c) Mean STS curves for each of 3 clusters are plotted with black solid line. The PCA-derived deviation from mean curve within each cluster is shown by dotted color lines. (d-f) PCA scree plots showing variance within each of 3 clusters.

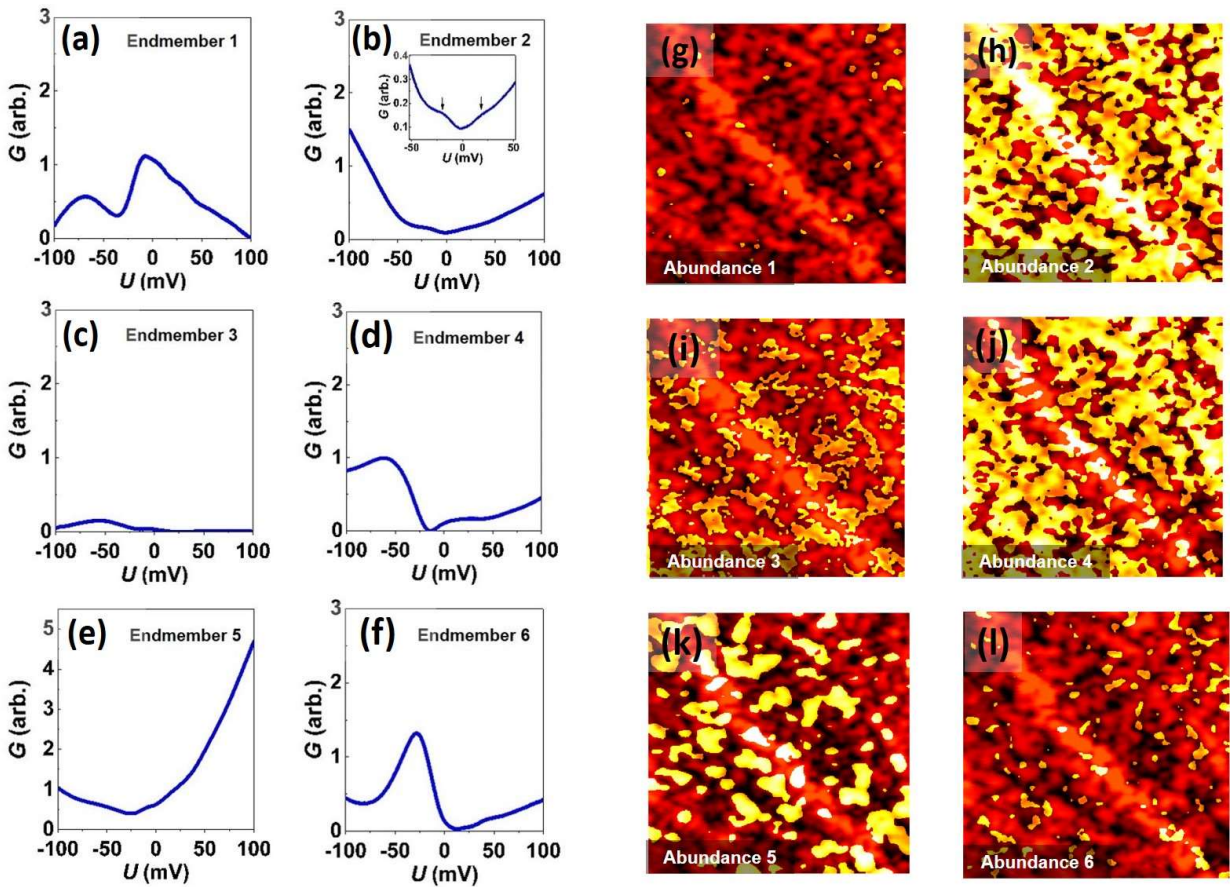

**FIGURE 5.** Results of Bayesian linear unmixing of STS dataset. (a-f) 6 Bayesian endmembers (see text for details). Inset in (b) zooms in the spectral gap features at the Fermi level. (g-l) Corresponding abundance maps (yellow) overlaid on the topographic image (red). The intensity contributions in the abundance maps below 0.2 were cut off for a better visualization (see SM for a full (0,1) intensity maps).